\documentclass{wiley-article}
\usepackage[numbers]{natbib}
\usepackage{siunitx}
\usepackage{lineno,hyperref}
\modulolinenumbers[5]
\usepackage{amsmath}
\usepackage{amssymb}

\usepackage{algorithm}
\usepackage{algpseudocode}
\usepackage{caption}
\usepackage{subcaption}
\usepackage{booktabs}
\usepackage{graphicx}
\newtheorem{thm}{Theorem}

\papertype{Research Article}
\title{Damping Identification of an Operational Offshore Wind Turbine using Kalman filter-based Subspace Identification.}

\author[1]{Aemilius A. W. van Vondelen}
\author[2]{Alexandros Iliopoulos}
\author[2]{Sachin T. Navalkar}
\author[1]{Daan C. van der Hoek}
\author[1]{Jan-Willem van Wingerden}

\affil[1]{Delft Center for Systems and Control, Delft University of Technology, Delft, 2628 CD, The Netherlands}
\affil[2]{Siemens Gamesa Renewable Energy, The Hague, 2595 BN, The Netherlands}

\corraddress{Aemilius A. W. van Vondelen, Delft Center for Systems and Control, Delft University of Technology, Delft, 2628 CD, The Netherlands}
\corremail{A.A.W.vanVondelen@tudelft.nl}

\fundinginfo{Not funded}

\runningauthor{van Vondelen et al.}

\begin{document}

\begin{frontmatter}
\maketitle

\begin{abstract}
Operational Modal Analysis (OMA) provides essential insights into the structural dynamics of an Offshore Wind Turbine (OWT). In these dynamics, damping is considered an especially important parameter as it governs the magnitude of the response at the natural frequencies. Violation of the stationary white noise excitation requirement of classical OMA algorithms has troubled the identification of operational OWTs due to harmonic excitation caused by rotor rotation. Recently, a novel algorithm was presented that mitigates harmonics by estimating a harmonic subsignal using a Kalman filter and orthogonally removing this signal from the response signal, after which the Stochastic Subspace Identification algorithm is used to identify the system. In this paper, the algorithm is tested on field data obtained from a multi-megawatt operational OWT using an economical sensor setup with two accelerometer levels. The first three tower bending modes could be distinguished, and, through the LQ-decomposition used in the algorithm, the identification results could be improved further by concatenating multiple datasets. A comparison against established harmonics-mitigating algorithms, Modified Least-squared Complex Exponential and PolyMAX, was done to validate the results.

% Please include a maximum of seven keywords
\end{abstract}
\end{frontmatter}

\section{Introduction}

Damping identification of operational vibrating structures is commonly achieved by employing Operational Modal Analysis (OMA). The main benefit of this practice is that does not require knowledge of the exciting force, which can be challenging to obtain for large industrial structures. A well-known limitation of this practice is the stationary white noise constraint that is placed on the excitation force. Although most excitations can be assumed stationary white noise, such as the traffic passing over a bridge or the wind blowing against it, several applications exist that are also excited by harmonic loading, such as Offshore Wind Turbines (OWTs). These harmonics originate from the rotation of the rotor and their presence renders the application of conventional OMA techniques nontrivial. For instance, structural modes might not be identified due to disturbance of the response signal caused by harmonic loading. Furthermore, harmonics in the response spectrum might be mistaken for structural modes, or inaccuracies might develop in the identification results due to merging of the vibration response data generated by harmonic loading and ambient loading \cite{mohanty2004operational}.

Several attempts have been made to develop methods that deal with harmonic presence in OMA. An elementary method involves recognising harmonic components in response spectra as zero-damped modes~\cite{mohanty2004operational}. This technique generally does not apply to OWTs as variations in rotor velocity over a measurement length cause non-stationary harmonic components which are non-zero damped. 

The Least-squares Complex Exponential algorithm (LSCE) is a classical time-domain OMA algorithm and was modified by Mohanty et al. (2004a) to incorporate harmonics in the identification steps \cite{mohanty2004operational,brown1979parameter}. This extension was added to other classical algorithms as well \cite{mohanty2004modified,mohanty2004modified2,mohanty2006modified}. The authors caution that exact knowledge of the harmonic frequency is required to use these techniques to obtain accurate results. In continuation, this methodology was also incorporated in a Stochastic Subspace Identification (SSI) approach by Dong~et~al.~(2014)~\cite{dong2014operational}.

Besides modified classical algorithms, some preprocessing techniques attempt to filter out harmonics. Time-synchronous averaging (TSA) is a popular technique in gearbox modal analysis \cite{crystal,combet2007automated}, where harmonics are generally stationary. For non-stationary harmonics, Cepstrum editing has been proposed as a more suitable technique \cite{randall1982cepstrum,randall2012new}. However, this method requires careful application and can affect the damping~\cite{manzato2014removing}. Both preprocessing approaches have been applied to OWTs with inconclusive results. In Manzato et al. (2014a) \cite{manzato2014removing}, the frequency content was affected after applying Cepstrum editing, but the TSA approach yielded good results. Contradicting, in Manzato et al. (2014b) \cite{manzato2014advanced}, the results were not satisfying with TSA, most likely due to non-stationary harmonics, and a slightly different application of Cepstrum yielded good results. 

Other techniques involve harmonic localisation utilising statistical indicators, such as the Probability Density Function (PDF) \cite{brincker2000indicator}, Kurtosis \cite{jacobsen2007using,agneni2012method} and Entropy \cite{agneni2012method}, whose performance depends on the peakedness of the distribution of the harmonic component.

One of the latest categories of algorithms is independent of the input spectrum. Transmissibility functions contain modal information unaffected by harmonic excitation \cite{devriendt2007use}. However, they are not yet optimal for this application, as important limitations are still present. For instance, several of the approaches require different loading conditions or many output sensors, making applications complicated \cite{devriendt2007use,devriendt2008identification,devriendt2010operational}. Other implementations require the acting forces to be uncorrelated \cite{yan2012operational,yan2015enhanced,yan2019transmissibility}, which can be problematic for identification with correlated harmonics induced by the turbine rotor \cite{weijtjens2014dealing}.

Recently, a novel approach called Kalman filter-based Stochastic Subspace Identification (KF-SSI) was developed using the classical SSI algorithm \cite{van1991subspace} that involves state reconstruction using a Kalman filter and subsequent orthogonal removal of the harmonic subsignal from the original signal \cite{gres2020kalman}. This method was reviewed together with other state-of-the-art OMA algorithms on applicability to operational OWTs in Van Vondelen et al. (2022) \cite{van2021damping} and was rated favourably due to its effective harmonic mitigating properties. In this paper, it is investigated whether this algorithm is suitable for application to OWTs by applying it to field data obtained from an offshore wind farm and comparing the results against two other harmonics-mitigating algorithms: the Modified Least-squares Complex Exponential and the PolyMAX algorithm. Moreover, an identification framework is developed that includes localisation of the harmonics, concatenation of multiple datasets to enhance accuracy, and automatic interpretation of the identification results.

For identification of OWTs, the installed sensors required by the IEC 61400 International Standard \cite{IEC} are usually not enough to perform modal analysis. Although it is possible to identify the first-order mode as the most deviation for this mode takes place in the tower top where the IEC 61400 sensors are installed, higher-order modes are much more difficult to identify \cite{vanderhoek2017}. Additional equipment is required to also capture higher-order modes \cite{manzato2014removing}, but this is often a costly endeavour. Ozbek et al. provide a comprehensive overview of different sensor setups for OMA \cite{ozbek2013operational}. In identifying the structural dynamics of a multi-megawatt OWT in this paper, we will attempt to identify higher-order modes in an economical setup, where only two accelerometer levels will be used. This, however, does not allow distinguishing mode shapes, as at least three accelerometers are required.

The main contribution of this paper is hence threefold: 1) We apply and compare different methods for determining the harmonic rotor frequencies from field data of OWTs and recommend the most suitable method. 2) We apply the Kalman filter-based Stochastic Subspace Identification framework to an industrial problem using limited instrumentation and identify the damping. 3) We compare the results of the KF-SSI algorithm against well-established OMA methods.

The next section introduces the field setup and describes the data acquisition. Section \ref{sec:framework} presents the identification framework. Here, different methods for harmonic localisation are compared, along with methods to improve identification results using KF-SSI. In Section \ref{sec:application}, the results are analysed after applying the KF-SSI identification framework to a 6 MW operation OWT. This paper is concluded in Section \ref{sec:conclusion}.

\section{Acquiring data from an operational wind farm}
\label{sec:field}
\begin{figure}
    \centering
    \includegraphics[width=0.6\textwidth]{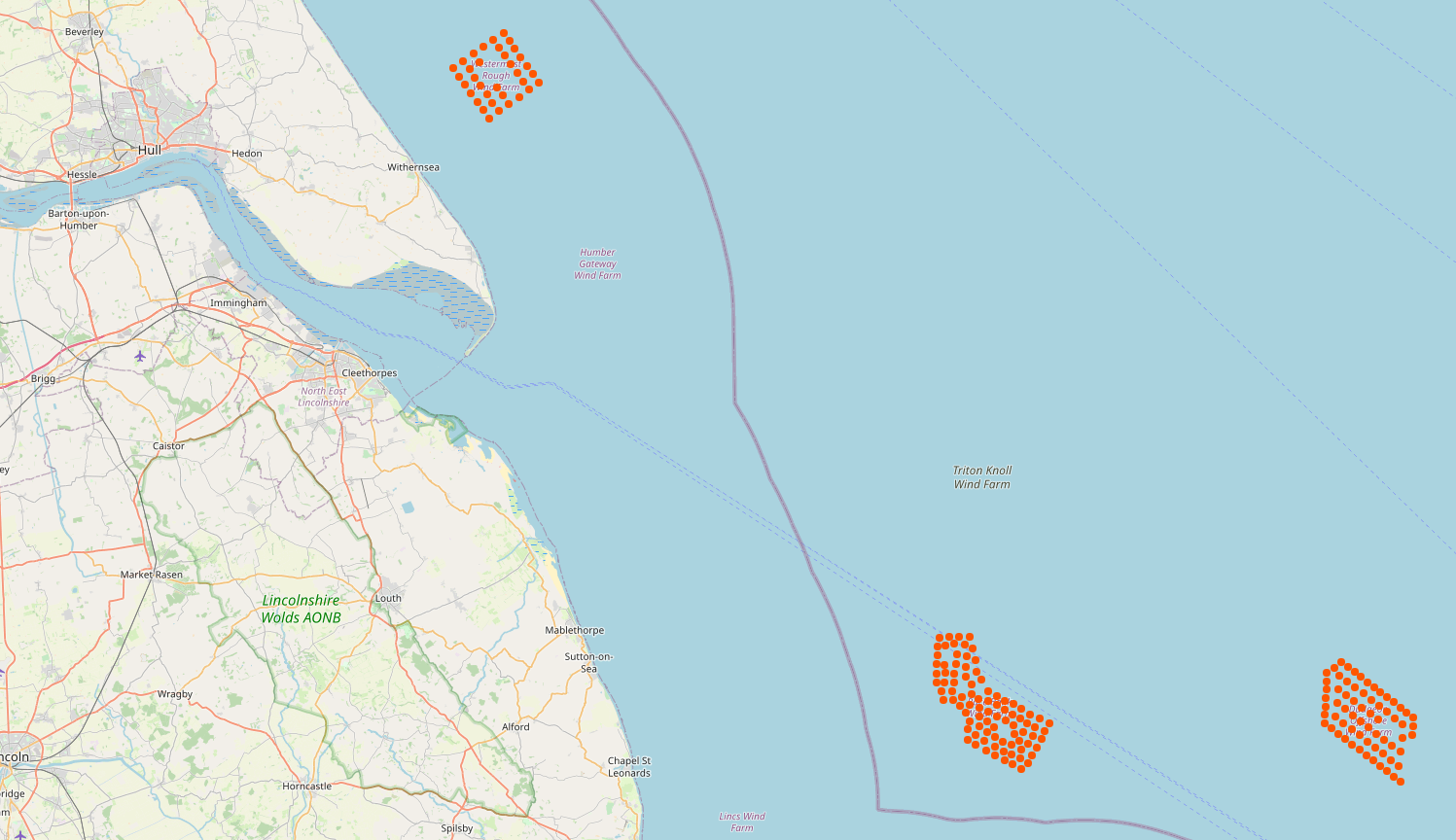}
    \caption{Lay out of the Dudgeon wind farm compared to other wind farms. From left to right: Westermost rough, Sheringham shoal, Dudgeon.}
    \label{fig:DudgeonWF}
\end{figure}

\begin{figure}
    \centering
    \includegraphics[clip, trim=0cm 13cm 8cm 4cm, width=0.45\textwidth]{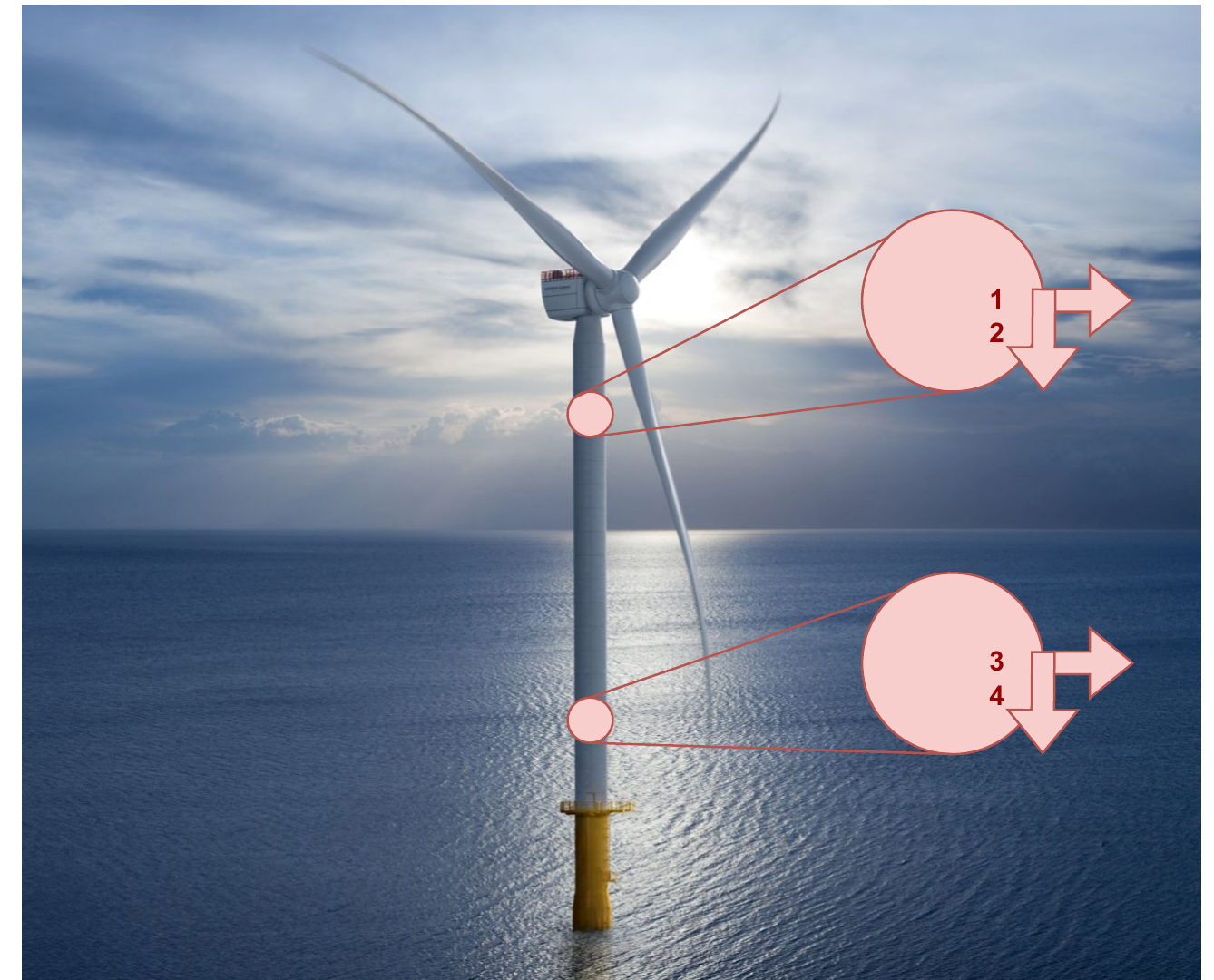}
    \caption{Locations of the four accelerometers mounted on the tower of the Dudgeon Offshore Wind Turbine.}
    \label{fig:SensLoc}
\end{figure}

\begin{figure}
    \centering
    \includegraphics[width=0.45\textwidth]{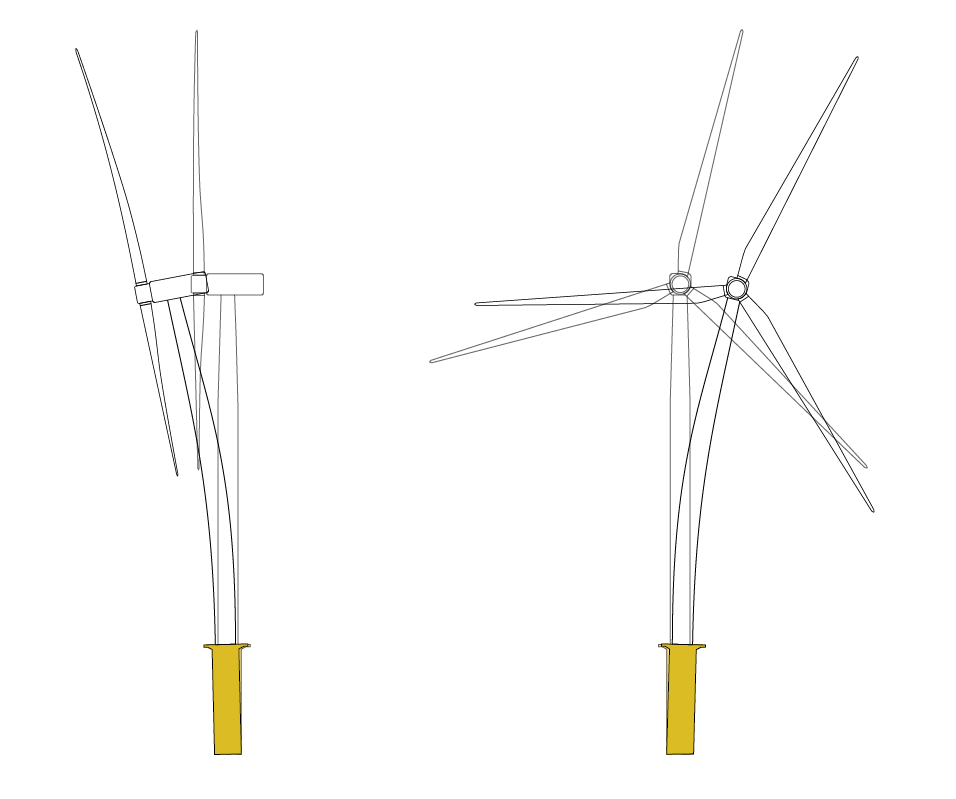}
    \caption{Illustration of the first Fore-Aft (left) and Side-Side (right) bending modes.}
    \label{fig:owtdirection}
\end{figure}

The field measurements were taken from a 6 MW Siemens Gamesa SWT-6.0-154 offshore wind turbine in the Dudgeon wind farm located in the United Kingdom 32 km off the coast of Cromer, a town in North Norfolk. The site has a water depth between 18 and 25 m, an average wind speed of 9.8 m/s and a mean wave height of 1.1 m. The turbines have monopile foundations, a rotor diameter of 154 m and a hub height of 110 m. The layout of the wind farm is illustrated in Figure \ref{fig:DudgeonWF}.
Several turbines in the Dudgeon wind farm are instrumented with two accelerometer levels with translation bi-axial accelerometers housed in a box mounted at the inside wall of the tower, offset of the vertical axis of the tower (See Figure \ref{fig:SensLoc} for an impression of the locations). The sensors are placed optimally for modal analysis but are constrained by the mounting options on the tower. Besides these sensors, the yaw and rotor velocity measurements were used for the identification procedure. The yaw signal measures the angular position of the rotor around the z-direction. This signal was used to transform the x and y directions of the accelerometers to the rotor coordinate system such that the transformed sensor outputs align with the direction of the Fore-Aft (FA) and Side-Side (SS) bending modes (see Figure \ref{fig:owtdirection} for an illustration of the first FA and SS modes). The generator rotational speed signal is measured by the turbine controller using an inductive sensor and interfaced to the data acquisition system. This signal was used to determine the location of the harmonics.
The datasets were sampled at a rate of 25 Hz, such that vibrations up to 12.5 Hz could be reconstructed. This frequency range captures all main modes and harmonics. For confidentiality reasons, all results have been normalised to fall between the values 0 and 1. The data was collected in 10-min batches. This dataset length can provide good estimates of the modal parameters. Better results can be obtained by concatenation of datasets, although this procedure is not trivial (see Section \ref{sec:analysemultiple}).

\section{Identification framework}
\label{sec:framework}
This section will expand on the different identification steps that are taken in identifying the damping of the 6 MW OWT. First, a method is selected for harmonic localisation (Section \ref{sec:localising}). Next, a theory is implemented for analysing multiple datasets, such that identification results can be improved (Section \ref{sec:analysemultiple}), followed by a description of an automated way to interpret the results (Section \ref{sec:interpreting}).

\subsection{Localising the Harmonics}
\label{sec:localising}
Removing harmonics using the KF-SSI algorithm requires knowledge of their frequency location. The Kurtosis indicator, as used in Jacobsen et al. (2007) \cite{jacobsen2007eliminating}, was initially suggested to distinguish harmonics from structural modes by Gre{\'s} et al. (2020) \cite{gres2020kalman}. Besides Kurtosis, other approaches will be considered in this section to determine the location of the harmonics, such as the Entropy approach, as used in Agneni et al. (2012) \cite{agneni2012method}. Finally, the rotor velocity method is considered as suggested in e.g. \cite{manzato2012review}. Note that the rotor velocity method only applies to systems where the rotating components causing the disturbing harmonics are measured (i.e., helicopter, OWT). The best-suited approach for an OWT will be selected to localise the harmonics.

\subsubsection{Harmonic Localisation using Statistical Indicators}
The first approach type is the use of statistical indicators, Kurtosis and Entropy, which use properties of the distribution of the response signal to determine the location of a harmonic.

The Kurtosis $\gamma$ can be computed at each frequency of the response signal by shifting a narrow bandpass filter over the entire frequency band of interest. This computation results in a value of $\gamma = 3$ for a signal that has a standard normal distribution, and a value of $\gamma \approx 1.5$ for harmonic signals that consist of sinusoids with zero mean $\mu$ and unit variance $\sigma^2$. Based on the difference in value, a distinction can be made between the two types of distributions.

Entropy attains a zero value if and only if the occurrence of a value is certain. Otherwise, it has a positive value. This property can be exploited to distinguish deterministic signals, such as harmonics, from stochastic processes. The Entropy value is non-maximum in deterministic components and maximum in regions of stochastic components.

\begin{figure}
    \centering
    \includegraphics[width=.6\textwidth]{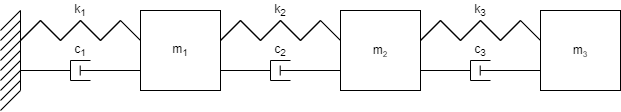}
    \caption{Illustration of the 3 degree-of-freedom system where $m_i, k_i$ and $c_i$ for $i=1,2,3$ are the mass, stiffness and damping, respectively.}
    \label{fig:3DOFfig}
\end{figure}

\begin{figure}
    \centering
    \includegraphics[clip, trim=0cm 11.5cm 0cm 11.5cm, width=\textwidth]{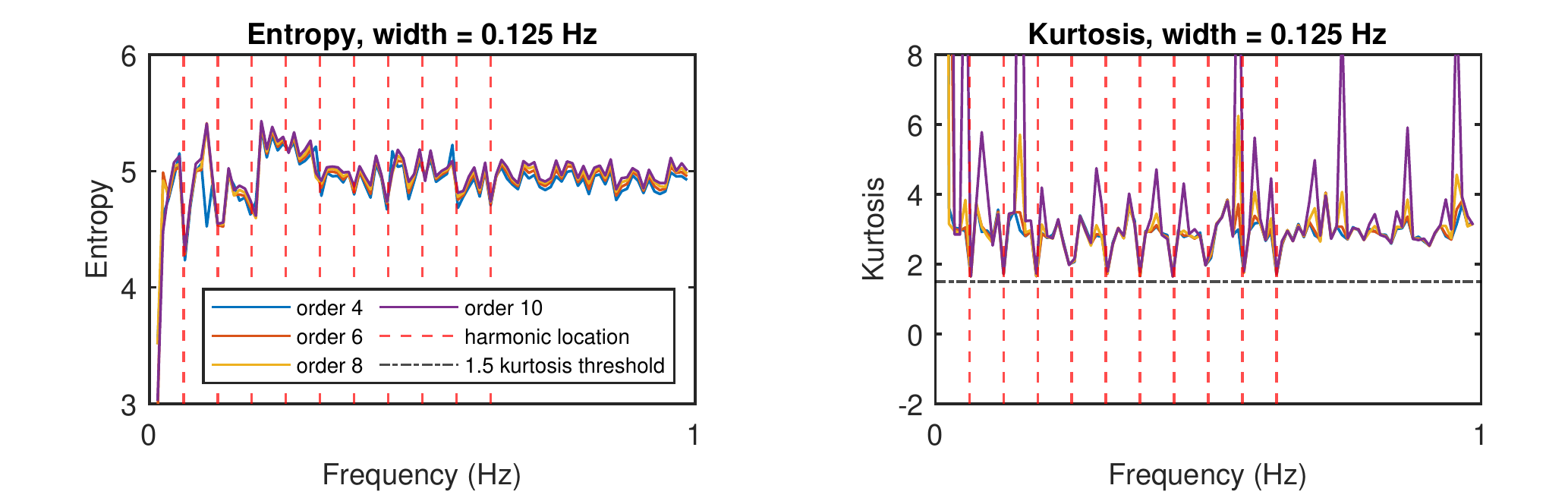}
    \caption{Comparison of the Kurtosis and Entropy statistical indicator for different filter orders on a simple 3 degree of freedom system with 10 harmonics.}
    \label{fig:3dofkurt}
\end{figure}

\begin{figure}
    \centering
    \includegraphics[clip, trim=0cm 11.5cm 0cm 11.5cm, width=\textwidth]{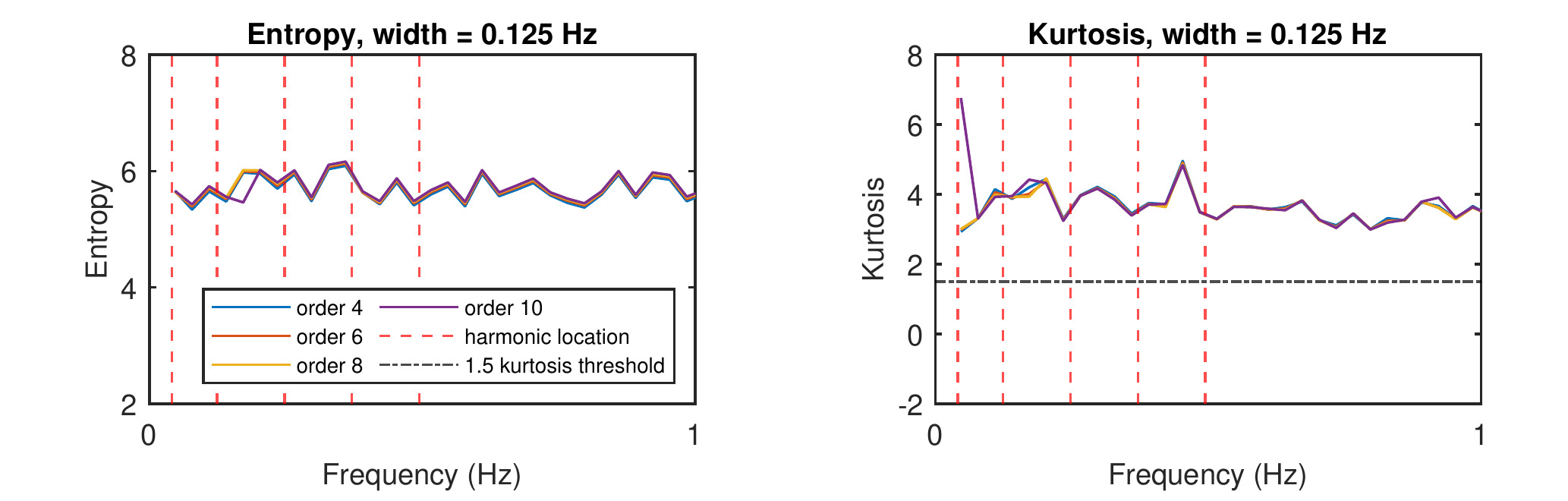}
    \caption{Comparison of the Kurtosis and Entropy statistical indicator for different filter orders on a field-measured dataset.}
    \label{fig:fdkurt}
\end{figure}

Both statistical indicators have been tested on simulation data of a simple 3 DoF system illustrated in Figure \ref{fig:3DOFfig} and field OWT data (Figure \ref{fig:3dofkurt}). Each signal is bandpass filtered by passing a Butterworth filter over the entire frequency range. The vertical dashed red lines indicate where the harmonics are located. The Kurtosis and Entropy statistical indicators are compared for different filter orders. It can be observed that the local minima in the Entropy plot indicate the harmonic locations. However, other local minima can be observed that do not indicate a harmonic, complicating localisation without prior knowledge. In the Kurtosis plot, the results are accurate for all harmonics, as the Kurtosis value attains 1.5 at each harmonic frequency.
Remarkably, both indicators show poor performance when applied to the field data as can be seen in Figure \ref{fig:fdkurt}. Here, the 1P, 3P, 6P, 9P and 12P harmonics are indicated by the red dashed lines. No consistent correspondence to the expected harmonics is visible in the local minima of the Entropy plots. Likewise for Kurtosis, no attainment of the 1.5 value can be observed at the expected locations of the harmonics.

Poor performance might be attributed to the broader harmonic peaks in the response spectra (see e.g. Figure \ref{fig:griddrop}) of field data from an OWT. The peaks are not as sharp as the peaks caused by the ideal harmonics in the response of the 3 DoF system, and therefore, a more comprehensive filter might be required to capture the entire harmonic signal. However, a wider filter could result in less accurate localisation of the harmonic and undermine its purpose. In a subsequent experiment, large filter widths were taken, which did not result in better localisation.

From this analysis, it can be concluded that the statistical indicators are not directly suitable for the localisation of harmonics in the identification problem of the OWT.

\subsubsection{Harmonic Localisation using the Rotor Velocity Signal}
In OWT measurements, the rotor velocity is generally available. This signal can be used for harmonic localisation but it is not constant over time. Therefore, the mean value over the entire sample was used as an estimate of the base frequency of the rotor and its harmonics. This method is expected to work best for approximately stationary rotational velocities, as the harmonic peak will be concentrated around a single frequency value. For less stationary rotational velocities, the peak will be less sharp, and estimation accuracy might deteriorate.

Field data from an OWT grid drop event is used to investigate whether this method provides accurate estimations. During a grid drop event, the power of the turbine drops to zero within a few milliseconds, which gives a massive thrust impulse to the OWT and sets up long-lasting tower oscillations. The speed reduction also causes a reduction in damping, leading to longer-lasting oscillations. The turbine then is in `idling' state, where no electricity is generated. Because the turbine tower is still vibrating from the operational state and is excited by wind and wave loading, the acceleration sensors provide the tower response from these loads only and no longer from the periodic loading caused by the rotor. 

\begin{figure}
    \centering
    \includegraphics[clip, trim=0cm 11.25cm 0cm 11.5cm, width=\textwidth]{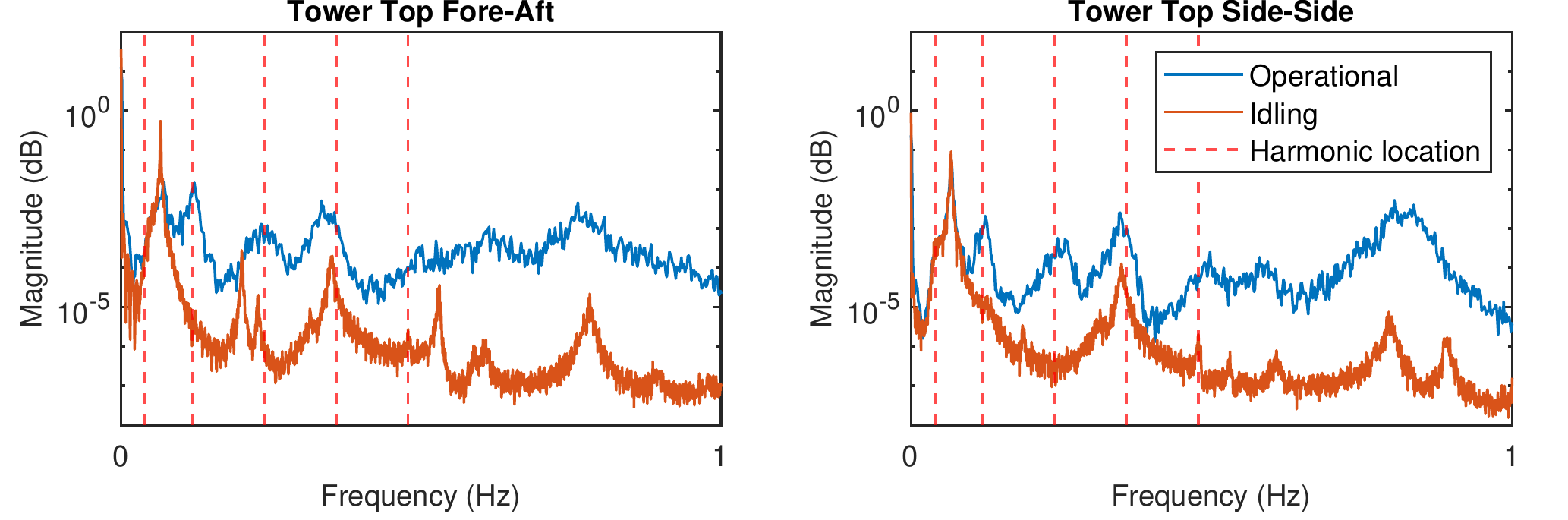}
    \caption{Comparison of the response spectra of the tower sensors using idling-only and operational data.}
    \label{fig:griddrop}
\end{figure}

When comparing the time series from both states, idling and operational, different response spectra are found. In Figure \ref{fig:griddrop}, the red vertical dashed lines indicate the harmonics 1P, 3P, 6P, 9P, and 12P estimated based on the rotational velocity. These lines agree with distinct peaks in the spectrum of the full time series, where the operational state is considered. When inspecting the response spectrum of the idling state, the indicated peaks have vanished, which suggests that they are associated with harmonics. Several modes that were clouded by the harmonics also become visible now.

\subsubsection{Conclusion}
Based on the above analyses of harmonic localisation methods, the rotor velocity approach is a suitable method for identifying the locations of the harmonics of an operational OWT. Therefore, this method is selected for the KF-SSI algorithm instead of the statistical methods.

\subsection{Analysing multiple datasets}
\label{sec:analysemultiple}
Improving identification results can be achieved by using more data. However, multiple distinct datasets cannot simply be assembled, as there will be a state mismatch between the final entry of the first data sequence and the first entry of the second data sequence, which could yield incorrect results. Especially for wind turbines, where the dynamics change with wind speed, it may be required to assemble multiple datasets from distinct periods, where the environmental and operational conditions were found to be similar. 

The KF-SSI algorithm can be implemented in a numerically efficient way such that computation time is reduced and numerical robustness is achieved using an LQ decomposition (\cite{gres2020kalman}, Section 3.4). This decomposition can also be exploited to effectively concatenate multiple datasets, because no state irregularity arises \cite{dohler2012fast,mastronardi2001fast}. A theorem is adapted from \cite{dohler2012fast} and presented below using notation consistent with (\cite{gres2020kalman}, Section 3.4): $Y_{\text{per}}$ is a hankel matrix containing data from the periodic subsignal, $Y_{\text{raw}}$ is a hankel matrix containing data from the original raw signal. The results obtained using this concatenation are hereafter referred to as obtained using Enhanced KF-SSI and compared against KF-SSI without concatenation in Section \ref{sec:mfdata}.
\begin{thm}
\label{thm2}
Consider the LQ decomposition of dataset 1
\begin{align}
\label{eq:stacked}
\begin{bmatrix}
Y_{\text{per}} \\
Y_{\text{raw}}
\end{bmatrix}_1
=
L_1 Q_1^T.
\end{align}
One can concatenate the next dataset 2 horizontally with dataset 1 as long as there is no state-discontinuity.
\begin{align}
\begin{bmatrix}
\begin{bmatrix}
Y_{\text{per}} \\
Y_{\text{raw}}
\end{bmatrix}_1
\begin{bmatrix}
Y_{\text{per}} \\
Y_{\text{raw}}
\end{bmatrix}_2
\end{bmatrix}
=
L_2 Q_2^T.
\end{align}
Alternatively, the two datasets can be concatenated irrespective of state-accordance, which yields the same $L$ matrix but a different $Q$ matrix:
\begin{align}
\begin{bmatrix}
L_1 &
\begin{bmatrix}
Y_{\text{per}} \\
Y_{\text{raw}}
\end{bmatrix}_2
\end{bmatrix}
=
L_2 Q_3^T.
\end{align}
\end{thm}
\begin{proof}[Proof of Theorem \ref{thm2}]
See Appendix \ref{sec:app}.
\end{proof}

\subsection{Interpreting Identification results}
\label{sec:interpreting}
To prevent the requirements of extensive manual analysis, a method was developed to interpret the stabilisation diagrams automatically. After identifying for a range of system orders, all unique poles within a tolerance that occur at least $n$ times are selected. Subsequently, the minimum order is found at which the maximum number of these unique poles is identified. This order is then selected as the optimal order and the modes identified at this order are used in the analysis. A pseudo-code is given below:
\begin{algorithm}
\caption{Automatic interpretation Stabilisation diagram}\label{euclid}
\begin{algorithmic}[1]
\For{order = 1, N}
\State{Run OMA algorithm with data $Y$}
\State{Stack row vector of identified damping $d$ in set $D$ and natural frequency $f$ in set $F$}
\EndFor
\State{Find all unique (within a tolerance) natural frequencies $F_{unique}$ in set $F$}
\For{order = 1, N}
\State{Check how many in set $F_{unique}$ exist in each row vector (order) of $F$}
\State{Select index of row vector and store in set $O$}
\EndFor
\State{Determine indices of row vectors containing most entries of $F_{unique}$} 
\State{Select lowest index and take $F(index)$ and $D(index)$ as final estimate result}
\end{algorithmic}
\end{algorithm}

\section{Application to an Operational Offshore Wind Turbine}
\label{sec:application}
The damping and natural frequency of a 6 MW OWT (Section \ref{sec:field}) are estimated in this section. There are only two accelerometer levels available at the tower top and the tower bottom, which is a minimal setup for higher mode identification. Mode shape estimation is not possible as a minimum of three accelerometer levels is required. Sufficient data is sampled at 25 Hz in ten minute lengths between wind speed range 5-26 m/s. In total, ten distinct datasets were used at each wind speed. 

The KF-SSI algorithm uses the conventional Kalman filter in its identification framework. However, numerical issues were encountered when implementing this filter for this application. Therefore, the authors used the Square-root Covariance filter as a more numerically robust alternative to the conventional Kalman filter \cite{verhaegen1986numerical}.

\subsection{Single field-measured dataset}
\begin{figure}
\centering
    \begin{subfigure}[t]{.52\textwidth}
        \centering
        \includegraphics[clip, trim=3cm 9cm 2cm 10cm, width=\textwidth]{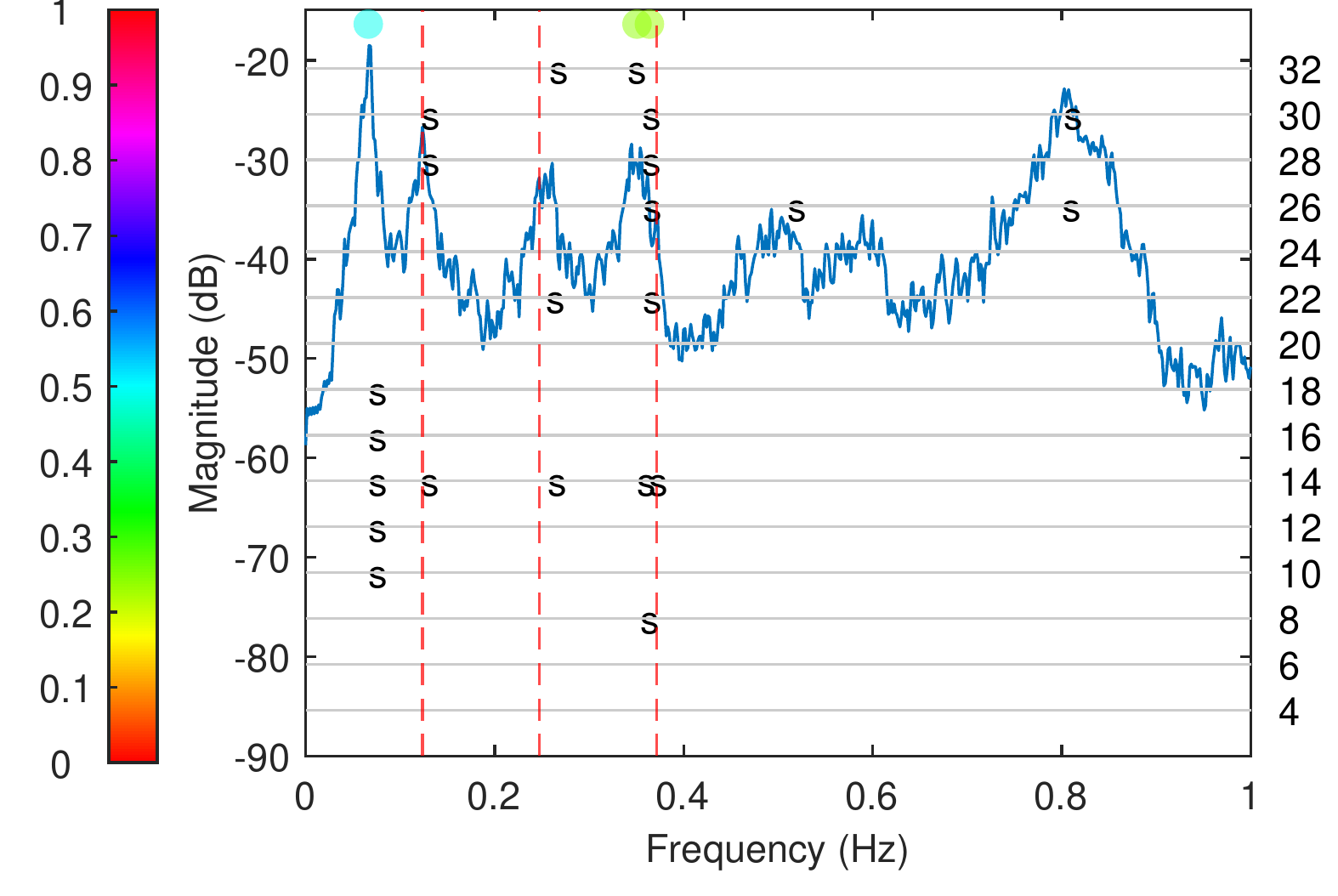}
        \caption{SSI.}
        \label{fig:SSkfssifd1}
    \end{subfigure}
\\
    \begin{subfigure}[t]{.49\textwidth}
        \centering
        \includegraphics[clip, trim=3cm 9cm 3cm 10cm, width=\textwidth]{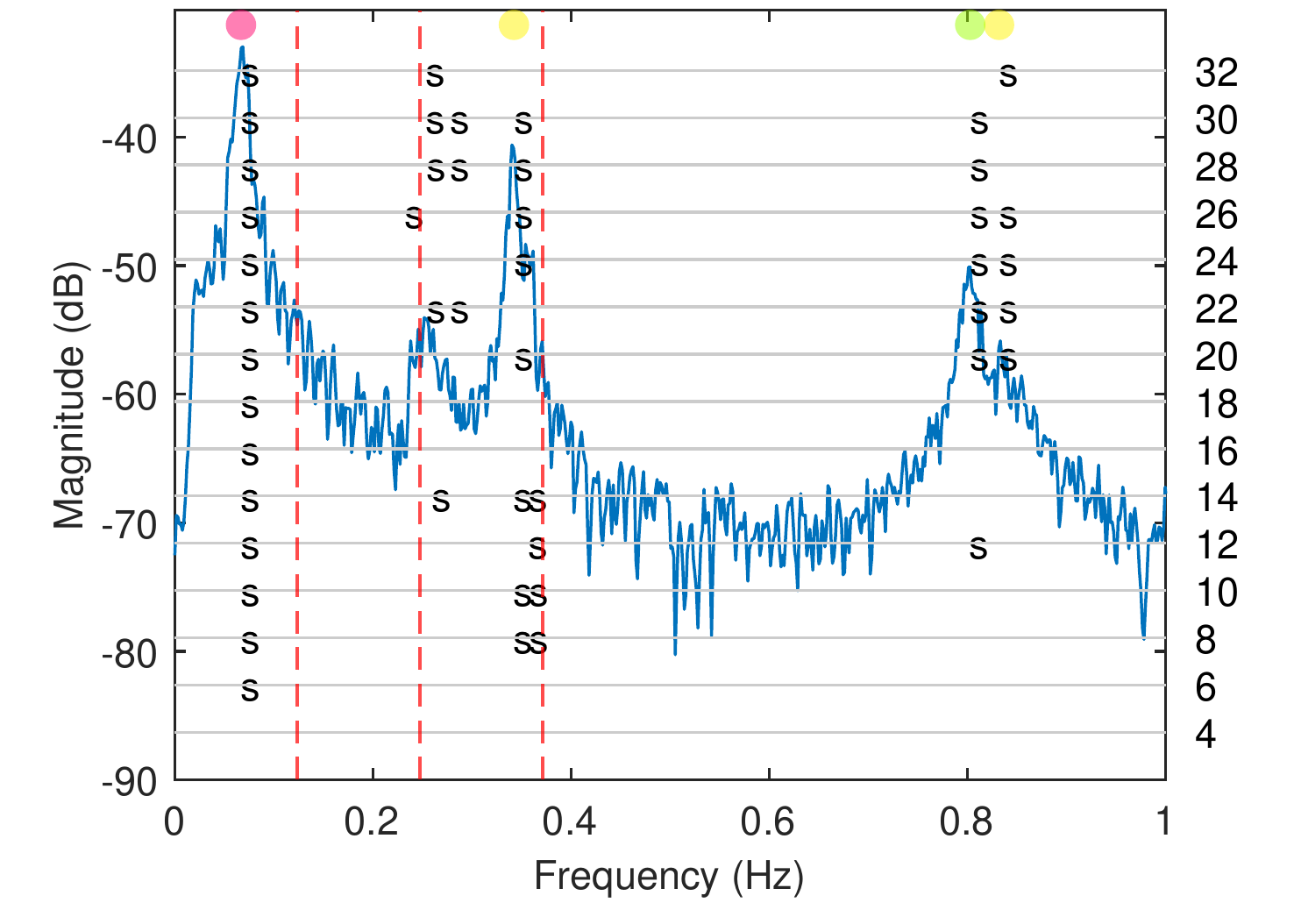}
        \caption{KF-SSI with 3P, 6P, and 9P orthogonally removed.}
        \label{fig:SSkfssifd5}
    \end{subfigure}
        \begin{subfigure}[t]{.49\textwidth}
        \centering
        \includegraphics[clip, trim=3cm 9cm 3cm 10cm, width=\textwidth]{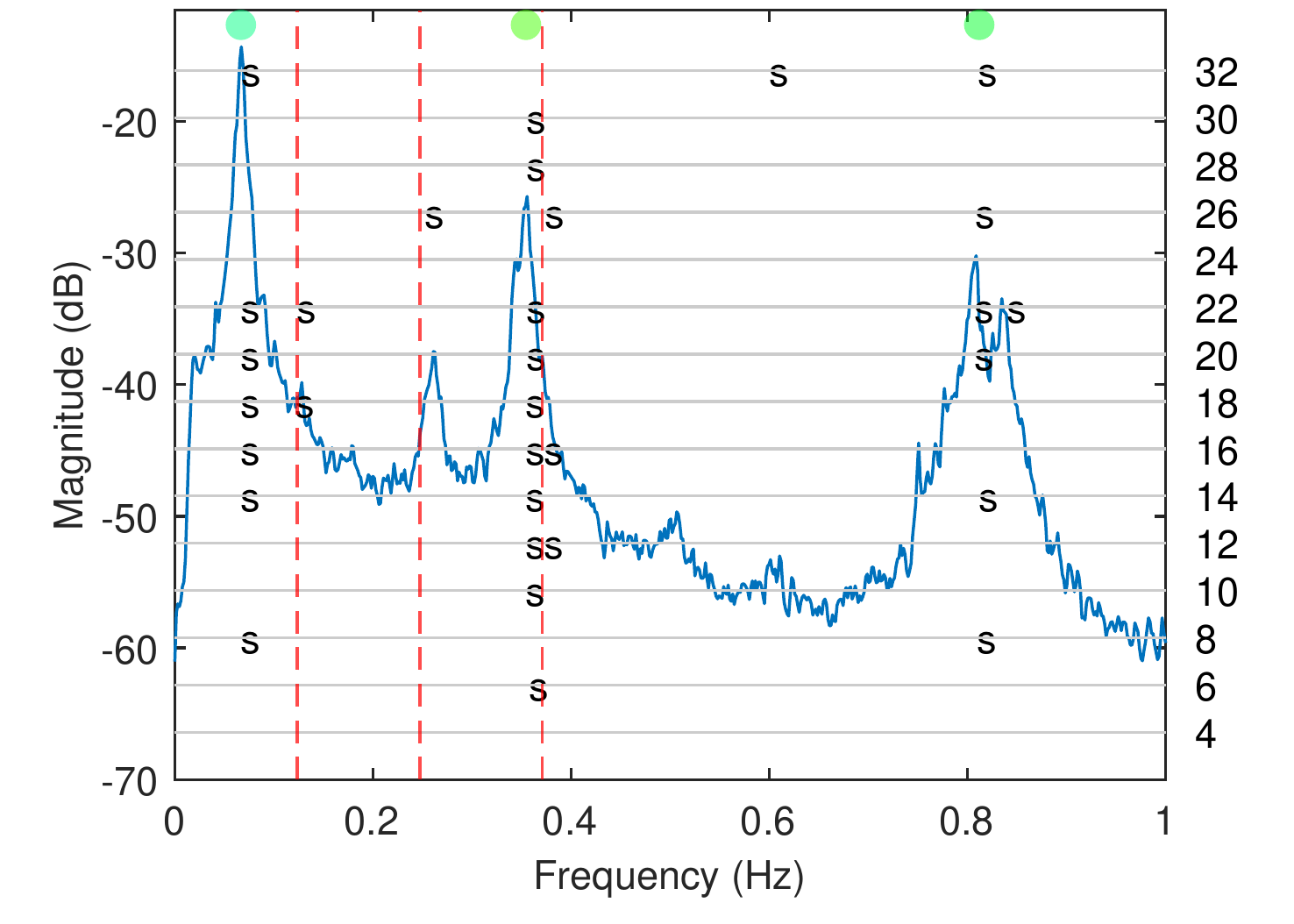}
        \caption{Enhanced KF-SSI using 10 datasets with 3P, 6P, and 9P orthogonally removed. Note that the spectrum presented here is illustrative as it cannot be reconstructed.}
        \label{fig:SSkfssifd2}
    \end{subfigure}
    \caption{Comparison of the stabilisation diagrams constructed from the results of the field data experiment using different settings of the KF-SSI algorithm for the SS direction. The red dashed lines indicate the harmonics that are orthogonally removed and the right axis indicates the identification order.}
    \label{fig:SSkfssifdtotal}
\end{figure}

First, the Enhanced KF-SSI framework is applied to a single dataset at an above-rated wind speed of 13 m/s for the Side-Side direction and compared against classical KF-SSI and SSI. The 3P, 6P and 9P harmonics are removed using the (Enhanced) KF-SSI algorithms. Consequently, stabilisation diagrams are generated for analysis and displayed in Figure \ref{fig:SSkfssifdtotal}.

The stabilization diagram provides the analyst with an overview of identified poles for a range of orders, such that an optimal order can be found in a heuristic approach. The power spectrum of the identified signal is plotted and the order is incremented by two with each identification cycle. If the algorithm finds a pole at an identification order that is within a user-defined tolerance of the previous order, an `S' is plotted at the identified natural frequency in the figure on the horizontal line belonging to the order, which is indicated on the right vertical axis. The damping is indicated by the coloured dots at the top of the figure and the harmonics by the red dashed lines.
An immediate difference can be observed when comparing the power spectra in the stabilisation diagrams of SSI and KF-SSI. Several peaks indicated by the red dashed lines have vanished, which are the harmonics 3P, 6P, and 9P. The 3P harmonic was reasonably spaced away from the first mode and the SSI algorithm was able to identify the first mode properly at several identification orders. However, the harmonic presence can still affect the estimated modal parameters \cite{gres2020kalman}. Removal of the 3P harmonic allows easier identification across multiple orders, as can be seen in the KF-SSI plot. 

After removing the 6P harmonic, there is still a peak visible in the spectrum. As no modes are expected to be found in that region, it might be an artefact of the removal of the 6P harmonic. Another reason might be that two modes were assigned to the 6P harmonic, while only one was removed. Strikingly, this phenomenon is observed in several datasets for different wind speeds. Further investigation is required to properly assess the origin of this remaining peak.

In the SSI stabilisation diagram, it can be seen that the 9P harmonic is right on top of the second mode, which is usually problematic for damping identification using classical OMA algorithms. Fortunately, the SSI algorithm yields an estimate. However, this estimate is likely contaminated by the damping of the 9P harmonic, due to its proximity and the fact that no additional mode was identified separating both damping values. Isolating and removing this harmonic using KF-SSI allows more accurate estimation of the second mode. This can be seen in the KF-SSI stabilisation diagram. This diagram shows clear results for the second mode. Also, due to the harmonic removal, estimates were found for the third bending mode around 0.8 Hz. 

All three modes were also found by the Enhanced KF-SSI. Note that unlike with regular KF-SSI the edited time signal \textit{cannot} be reconstructed from the Hankel matrices of the LQ decomposition due to the different $Q$ matrix (see Theorem \ref{thm2}). Hence, for illustrative purposes, the power spectrum given in Figure \ref{fig:SSkfssifd2}, normally constructed from the edited time signal with original KF-SSI, was created by averaging the individual power spectra obtained from each dataset after separate LQ decompositions. The estimated modal parameters are also given in Table \ref{tab:kfssifd1}. A notable difference is found between the estimates of KF-SSI and Enhanced KF-SSI. This difference demonstrates the desire for a method that allows the assembly of multiple datasets, such that estimation variance can be reduced. The next section will cover a more extensive comparison between KF-SSI and Enhanced KF-SSI for multiple datasets.
\begin{table}
    \caption{Comparison of the modal parameters obtained in the field data experiment for KF-SSI.}
       \begin{threeparttable}
        \begin{tabular}{lcccccc}
        \headrow 
       & \multicolumn{3}{c}{\textbf{Natural frequency (Hz)}}  & \multicolumn{3}{c}{\textbf{Damping $\delta$ (\%)}}\\
        \headrow
        \textbf{Mode}  & Classical & KF-SSI & Enhanced & Classical & KF-SSI & Enhanced\\
        \hline
        1              & .0663      & .0668 & .0667  & .500      & .943   & .431\\
        2              & .351      & .342 & .354   & .254      & .170   & .297\\
        3              & -         & .803 & .812   & -      & .243   & .373\\
       \hline
        \end{tabular}
\begin{tablenotes}
\item Abbreviations: Classical, Classical Stochastic Subspace Identification; KF-SSI, Kalman filter-based Stochastic Subspace Identification; Enhanced, Enhanced Kalman filter-based Stochastic Subspace Identification.
\end{tablenotes}
\end{threeparttable}
\label{tab:kfssifd1}
\end{table}

\subsection{Multiple field-measured datasets}
\label{sec:mfdata}
\begin{figure}
    \centering
    \includegraphics[clip, trim=0cm 9cm 0cm 8.5cm, width = \textwidth]{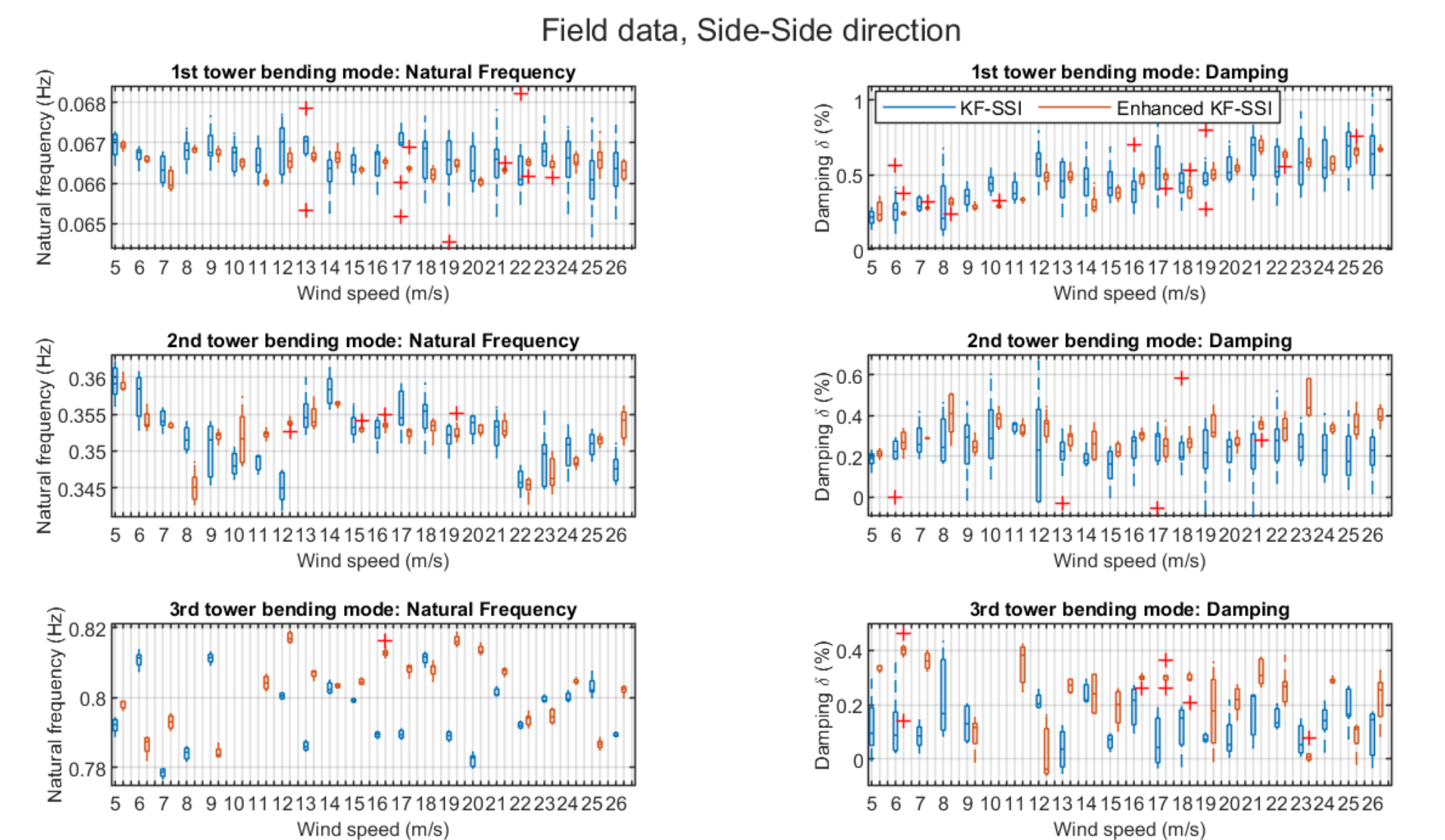}
    \caption{Comparison of Enhanced KF-SSI against original KF-SSI.}
    \label{fig:enhancedKFSSI}
\end{figure}

This section presents the results obtained from multiple datasets. Box plots are constructed for the Side-Side direction for both the KF-SSI algorithm and Enhanced KF-SSI algorithm using Leave-One-Out methodology and are displayed in Figure \ref{fig:enhancedKFSSI}. A notable reduction in the spread is visible for the enhanced version compared to the original version, which improves identification precision strongly. The first mode is identified persistently by both algorithms. There is a low variance for the estimated natural frequency and a more extensive variance for the damping estimates. Furthermore, there is an increasing trend visible for the damping of the first mode at higher wind speeds, which can be attributed to the increasing effect of aerodynamic damping for higher wind speeds. For the second mode, the spread is more prominent with the original algorithm at the natural frequencies and damping values. The damping value for the second mode is approximately constant across all wind speeds. Careful manual analysis could improve the results even more, but it is time-consuming as 220 datasets need to be interpreted and processed. The third mode shows an even larger spread, also for the enhanced algorithm. However, this is expected as higher-order modes are more difficult to estimate due to lower energy. Again for the third mode, the damping value remains approximately constant across all wind speeds.

\subsection{Comparison against other algorithms}
To validate the identification precision, the results obtained using the Enhanced KF-SSI algorithm are compared against two other algorithms. The first algorithm is called PolyMAX and is a well-established frequency-domain OMA algorithm. It is favoured due to its fast convergence and clear stabilisation diagrams \cite{peeters2005polymax}. However, it does not employ harmonics-mitigating steps in its identification procedure, and therefore careful selection is required, as harmonics might be identified as false modes. The second algorithm for comparison is the Modified Least-squares Complex Exponential (LSCE) algorithm \cite{mohanty2004operational}. This time-domain algorithm incorporates the harmonics in the identification steps and uses the same a priori information as the Enhanced KF-SSI algorithm based on the rotational velocity.

\begin{figure}[t]
    \centering 
    \includegraphics[clip, trim=0cm 9cm 0cm 8.5cm, width = \textwidth]{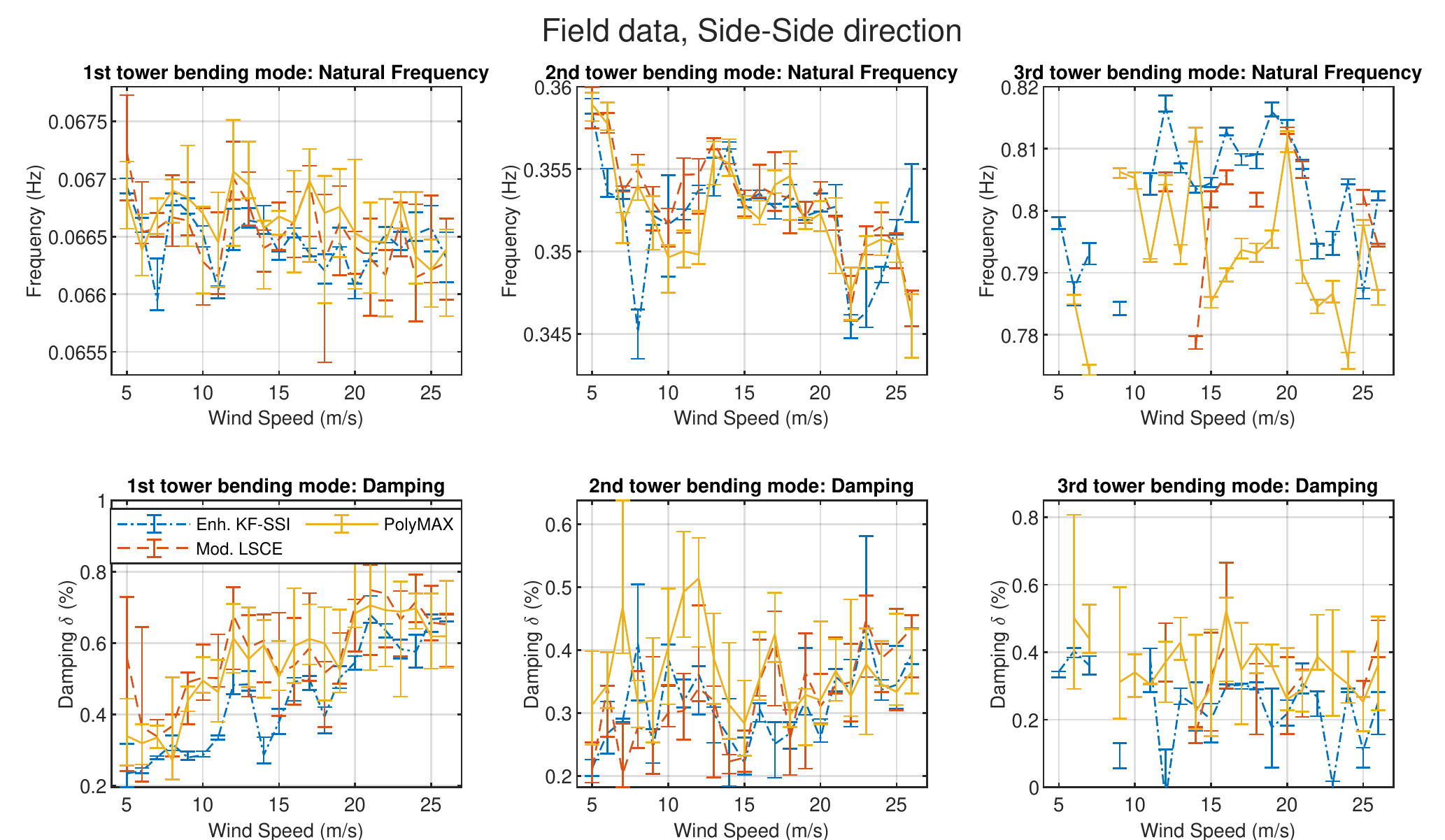}
    \caption{Comparison of Enhanced KF-SSI against PolyMAX and Modified LSCE.}
    \label{fig:compalgo}
\end{figure}

A comparison is made of the median, first and third quantiles of each algorithm on the same datasets as in the previous section. The resulting box plots are shown in Figure \ref{fig:compalgo}, and the median values are displayed in Tables \ref{tab:natfreqcomp} and \ref{tab:dampcomp}. Here, the small spread of the Enhanced KF-SSI compared to other algorithms is especially notable, and the median values correspond well with the other algorithms. More importantly, the overall trend is correctly captured by Enhanced KF-SSI. Another observation is that Enhanced KF-SSI shows persistency in identification for higher-order modes, whereas other algorithms fail to estimate at some wind speeds.

\begin{table}[hb]
    \caption{Comparison of the median natural frequency estimates of the Enhanced KF-SSI, PolyMAX, and Modified LSCE algorithms at different wind speeds.}
    \begin{threeparttable}
    \resizebox{\columnwidth}{!}{%
    \begin{tabular}{llccccccccccccccccccccccc}
    \headrow 
     \textbf{Mode} & \textbf{Algorithm} & 5 & 6 & 7 & 8 & 9 & 10 & 11 & 12 & 13 & 14 & 15 & 16 & 17 & 18 & 19 & 20 & 21 & 22 & 23 & 24 & 25 & 26 \\
     \hline
First & Enhanced KF-SSI & .0669 & .0666 & .0659 & .0668 & .0667 & .0665 & .0660 & .0665 & .0666 & .0666 & .0664 & .0665 & .0664 & .0662 & .0664 & .0661 & .0663 & .0665 & .0664 & .0665 & .0666 & .0663 \\
& PolyMAX & .0668 & .0664 & .0667 & .0669 & .0668 & .0667 & .0665 & .0671 & .0669 & .0666 & .0667 & .0666 & .0670 & .0667 & .0668 & .0665 & .0665 & .0665 & .0668 & .0663 & .0662 & .0664 \\
& Modified LSCE & .0672 & .0665 & .0666 & .0667 & .0666 & .0663 & .0662 & .0670 & .0668 & .0664 & .0665 & .0666 & .0670 & .0663 & .0666 & .0664 & .0663 & .0662 & .0666 & .0661 & .0662 & .0663 \\
     \hline
Second & Enhanced KF-SSI & .358 & .354 & .353 & .345 & .352 & .352 & .352 & .354 & .354 & .356 & .353 & .354 & .353 & .353 & .352 & .352 & .353 & .345 & .346 & .348 & .352 & .354 \\
& PolyMAX & .359 & .358 & .352 & .354 & .352 & .350 & .350 & .350 & .356 & .355 & .353 & .352 & .354 & .355 & .352 & .352 & .350 & .347 & .350 & .351 & .350 & .346 \\
& Modified LSCE & .358 & .358 & .354 & .355 & .353 & .352 & .355 & .355 & .357 & .355 & .353 & .354 & .353 & .353 & .352 & .354 & .352 & .347 & .351 & .352 & .350 & .347 \\
\hline
Third & Enhanced KF-SSI & .799 & .788 & .793 & - & .784 &  -    & .804 & .817 & .807 & .803 & .804 & .813 & .809 & .809 & .816 & .813 & .808 & .794 & .795 & .804 & .787 & .803 \\
& PolyMAX &  -    & .786 & .774 & - & .806 & .805 & .792 & .805 & .793 & .813 & .785 & .790 & .794 & .793 & .796 & .811 & .790 & .785 & .787 & .776 & .798 & .787 \\
& Modified LSCE &  -    &    -  &    -  & - &    -  &  -    &   -   & .804 &   -   & .778 & .802 & .806 &   -   & .802 &  -    & .813 & .807 &   -   &    -  & -     & .803 & .794 \\
    \hline
    \end{tabular}
    }
    \begin{tablenotes}
\item Abbreviations: KF-SSI, Kalman filter-based Stochastic Subspace Identification; LSCE, Least-Squares Complex Exponential.
\end{tablenotes}
\end{threeparttable}
    \label{tab:natfreqcomp}
\end{table}
\begin{table}[ht]
    \caption{Comparison of the median damping estimates of the Enhanced KF-SSI, PolyMAX, and Modified LSCE algorithms at different wind speeds.}
    \begin{threeparttable}
    \resizebox{\columnwidth}{!}{%
    \begin{tabular}{llccccccccccccccccccccccc}
    \headrow
     \textbf{Mode} & \textbf{Algorithm} & 5 & 6 & 7 & 8 & 9 & 10 & 11 & 12 & 13 & 14 & 15 & 16 & 17 & 18 & 19 & 20 & 21 & 22 & 23 & 24 & 25 & 26 \\
     \hline
First & Enhanced KF-SSI & 0.234 & 0.241 & 0.279 & 0.316 & 0.279 & 0.285 & 0.335 & 0.483  & 0.484 & 0.286 & 0.379 & 0.491 & 0.498 & 0.393 & 0.502 & 0.545 & 0.679 & 0.635 & 0.583 & 0.575 & 0.666 & 0.670 \\
& PolyMAX & 0.339 & 0.319 & 0.334 & 0.271 & 0.438 & 0.495 & 0.463 & 0.614  & 0.555 & 0.594 & 0.506 & 0.591 & 0.611 & 0.599 & 0.525 & 0.683 & 0.705 & 0.692 & 0.688 & 0.697 & 0.618 & 0.619 \\
& Modified LSCE & 0.561 & 0.364 & 0.338 & 0.367 & 0.489 & 0.502 & 0.501 & 0.677  & 0.586 & 0.607 & 0.515 & 0.537 & 0.585 & 0.516 & 0.551 & 0.702 & 0.748 & 0.739 & 0.666 & 0.716 & 0.657 & 0.651 \\
     \hline
Second & Enhanced KF-SSI & 0.213 & 0.267 & 0.288 & 0.411 & 0.243 & 0.386 & 0.318 & 0.361  & 0.298 & 0.261 & 0.221 & 0.306 & 0.251 & 0.267 & 0.317 & 0.260 & 0.357 & 0.338 & 0.438 & 0.337 & 0.345 & 0.393 \\
& PolyMAX & 0.313 & 0.345 & 0.468 & 0.315 & 0.321 & 0.405 & 0.492 & 0.514  & 0.386 & 0.313 & 0.284 & 0.356 & 0.425 & 0.297 & 0.329 & 0.321 & 0.367 & 0.327 & 0.377 & 0.345 & 0.333 & 0.365 \\
& Modified LSCE & 0.210 & 0.341 & 0.203 & 0.278 & 0.267 & 0.300 & 0.303 & 0.339  & 0.320 & 0.223 & 0.229 & 0.351 & 0.411 & 0.258 & 0.364 & 0.311 & 0.343 & 0.349 & 0.447 & 0.387 & 0.408 & 0.435 \\
\hline
Third & Enhanced KF-SSI & 0.343 & 0.403 & 0.361 &    -   & 0.116 &     -  & 0.382 & -0.039 & 0.271 & 0.240 & 0.201 & 0.299 & 0.303 & 0.293 & 0.177 & 0.218 & 0.307 & 0.267 & 0.008 & 0.285 & 0.110 & 0.255 \\
& PolyMAX &    -   & 0.501 & 0.440 &     -  & 0.311 & 0.340 & 0.305 & 0.369  & 0.429 & 0.225 & 0.297 & 0.522 & 0.347 & 0.417 & 0.359 & 0.261 & 0.288 & 0.387 & 0.342 & 0.306 & 0.253 & 0.364 \\
& Modified LSCE &    -    &    -   &    -   &     -  &    -   &    -   &    -   & 0.429  &   -    & 0.159 & 0.332 & 0.426 &    -   & 0.275 &    -   & 0.272 & 0.327 &    -   &    -   &     -  & 0.291 & 0.441 \\
    \hline
    \end{tabular}
    }
    \begin{tablenotes}
\item Abbreviations: KF-SSI, Kalman filter-based Stochastic Subspace Identification; LSCE, Least-Squares Complex Exponential.
\end{tablenotes}
\end{threeparttable}
   \label{tab:dampcomp}
\end{table}

\section{Conclusions}
\label{sec:conclusion}
This study used the recently proposed Kalman filter-based Stochastic Subspace Identification (KF-SSI) algorithm in an identification framework that was used to identify the damping from field data obtained from an operational Offshore Wind Turbine (OWT). The KF-SSI algorithm mitigates harmonic disturbances and is therefore attractive for application to OWTs that are in operating condition.

Moreover, a comparison was made of methods that could indicate the harmonic locations, which is required a priori knowledge for KF-SSI. It was found that the statistical methods Kurtosis and Entropy were not able to identify the frequencies at which the harmonics are present from the field data. However, the mean value of the rotor velocity signal could effectively be used as estimate of these frequencies in the identification framework.

Furthermore, the authors enhanced the algorithm such that long or multiple datasets can be used that have been measured at different moments in time. This allows significant enhancements of the estimate precision through the use of more data. By subsequently applying the Leave-One-Out methodology, the median can be taken as a representative modal estimate. 

Excellent results were found for the damping and frequency of the first three tower bending modes using an economically attractive setup of only two accelerometer levels. More analyses can be done when more accelerometers are installed on the turbine, such as mode shape estimation. Finally, the results were compared against the established PolyMAX and Modified Least-squares Complex Exponential methods from which it could be deduced that the Enhanced KF-SSI method correctly captured the overall damping trend throughout the operational wind speed range.

\section*{Conflict of interest}
The authors declare that they have no conflict of interest.

\newpage
\appendix
\section{Proof of Theorem \ref{thm2}}
\label{sec:app}
\noindent The LQ decomposition of batch 1 is given as follows:
\begin{align}
    \begin{bmatrix} Y_{\text{per}} \\ Y_{\text{raw}} \end{bmatrix}_1
    =
    L_1 Q_1
    =
    \begin{bmatrix} L_{111} & 0 \\ L_{121} & L_{122} \end{bmatrix}
    \begin{bmatrix} Q_{11} \\ Q_{12} \end{bmatrix}.
    \label{eq:eq1}
\end{align}
\noindent Now consider the LQ decomposition of the concatenated batch 1 and batch 2:
\begin{align}
    \begin{bmatrix}
    \begin{bmatrix} Y_{\text{per}} \\ Y_{\text{raw}}\end{bmatrix}_1 & \begin{bmatrix} Y_{\text{per}} \\
     Y_{\text{raw}}
    \end{bmatrix}_2
    \end{bmatrix}
    =
    L_2 Q_2
    =
    \begin{bmatrix}
    L_{211} & 0 \\
    L_{221} & L_{222}
    \end{bmatrix}
    \begin{bmatrix}
    Q_{21} \\ Q_{22}
    \end{bmatrix}.
\end{align}
\noindent Next, consider the LQ decomposition of batch 1 and batch 2 using instead the $L_1$ of the LQ decomposition of batch 1 data (Equation \ref{eq:eq1}) as follows:
\begin{align}
    \begin{bmatrix}
    L_1 & \begin{bmatrix}
    Y_{\text{per}} \\ Y_{\text{raw}}
    \end{bmatrix}_2
    \end{bmatrix}
    =
    L_2 Q_3
    =
    \begin{bmatrix}
    L_{211} & 0 \\
    L_{221} & L_{222}
    \end{bmatrix}
    \begin{bmatrix}
    Q_{31} \\ Q_{32}
    \end{bmatrix}.
\end{align}
\noindent Observe that the $L_2$ matrix will be the same for both approaches, but the orthogonal Q matrix will be different. This can be verified by calculating the squares of all LQ decompositions:
\begin{align}
\notag
    L_1 Q_1 Q_1^T L_1^T &= L_1 L_1^T = \begin{bmatrix}
    L_{111} & 0 \\ L_{121} & L_{122}
    \end{bmatrix}
    \begin{bmatrix}
    L_{111} & 0 \\ L_{121} & L_{122}
    \end{bmatrix}^T \\ \notag
    &=
    \begin{bmatrix}
    L_{111} L_{111}^T & L_{111} [L_{121}\;\;L_{122}]^T \\ L_{111}^T [L_{121}\;\;L_{122}] & [L_{121}\;\;L_{122}] [L_{121}\;\;L_{122}]^T
    \end{bmatrix}
    \\ \label{eq:res3}
    &=
    \begin{bmatrix}
    Y_{\text{per}} \\ Y_{\text{raw}}
    \end{bmatrix}_1
    \begin{bmatrix}
    Y_{\text{per}} \\ Y_{\text{raw}}
    \end{bmatrix}_1^T
    =
    \begin{bmatrix}
    Y_{\text{per}_1} Y_{\text{per}_1}^T & Y_{\text{per}_1} Y_{\text{raw}_1}^T \\
    Y_{\text{raw}_1} Y_{\text{per}_1}^T & Y_{\text{raw}_1} Y_{\text{raw}_1}^T
    \end{bmatrix}, \\[10pt]
\notag
    L_2 Q_2 Q_2^T L_2^T &= L_2 L_2^T = \begin{bmatrix}
    \begin{bmatrix} Y_{\text{per}} \\ Y_{\text{raw}}\end{bmatrix}_1 & \begin{bmatrix} Y_{\text{per}} \\
     Y_{\text{raw}}
    \end{bmatrix}_2
    \end{bmatrix}
    \begin{bmatrix}
    \begin{bmatrix} Y_{\text{per}} \\ Y_{\text{raw}}\end{bmatrix}_1 & \begin{bmatrix} Y_{\text{per}} \\
     Y_{\text{raw}}
    \end{bmatrix}_2
    \end{bmatrix}^T
    \\ \label{eq:res1}
    &=
    \begin{bmatrix}
    Y_{\text{per}_1} Y_{\text{per}_1}^T + Y_{\text{per}_2} Y_{\text{per}_2}^T & Y_{\text{per}_1} Y_{\text{raw}_1}^T + Y_{\text{per}_2} Y_{\text{raw}_2}^T \\
    Y_{\text{raw}_1} Y_{\text{per}_1}^T + Y_{\text{raw}_2} Y_{\text{per}_2}^T & Y_{\text{raw}_1} Y_{\text{raw}_1}^T + Y_{\text{raw}_2} Y_{\text{raw}_2}^T
    \end{bmatrix}, \\[10pt]
\notag
    L_2 Q_3 Q_3^T L_2^T &= L_2 L_2^T = \begin{bmatrix}
    L_1 & \begin{bmatrix}
    Y_{\text{per}} \\ Y_{\text{raw}}
    \end{bmatrix}_2
    \end{bmatrix}
    \begin{bmatrix}
    L_1 & \begin{bmatrix}
    Y_{\text{per}} \\ Y_{\text{raw}}
    \end{bmatrix}_2
    \end{bmatrix}^T \\ \notag
    &=
    \begin{bmatrix}
    L_{111} & 0 & Y_{\text{per}_2} \\ L_{121} & L_{122} & Y_{\text{raw}_2}
    \end{bmatrix}
    \begin{bmatrix}
    L_{111} & 0 & Y_{\text{per}_2} \\ L_{121} & L_{122}& Y_{\text{raw}_2}
    \end{bmatrix}^T
    \\ \label{eq:res2}
    &=
    \begin{bmatrix}
    L_{111} L_{111}^T + Y_{\text{per}_2} Y_{\text{per}_2}^T & L_{111} [L_{121}\;\;L_{122}]^T + Y_{\text{per}_2} Y_{\text{raw}_2}^T \\
    L_{111}^T [L_{121}\;\;L_{122}] + Y_{\text{raw}_2} Y_{\text{per}_2}^T & [L_{121}\;\;L_{122}] [L_{121}\;\;L_{122}]^T + Y_{\text{raw}_2} Y_{\text{raw}_2}^T
    \end{bmatrix}.
\end{align}
The result of Equation \ref{eq:res3} can now be substituted in the result of Equation \ref{eq:res2} and will yield the same result as Equation~\ref{eq:res1}.

\printendnotes

% Submissions are not required to reflect the precise reference formatting of the journal (use of italics, bold etc.), however it is important that all key elements of each reference are included.
\bibliography{sample}

\end{document}